\documentclass[twocolumn,aps]{revtex4}
\usepackage{graphicx}%
\usepackage{dcolumn}
\usepackage{amsmath}
\begin{document}

\title{The dynamics of apparent horizons in Robinson-Trautman spacetimes}

\author{H. P. de Oliveira}
\email{oliveira@dft.if.uerj.br}
\affiliation{{\it Universidade do Estado do Rio de Janeiro }\\
{\it Instituto de F\'{\i}sica - Departamento de F\'{\i}sica Te\'orica}\\
{\it Cep 20550-013. Rio de Janeiro, RJ, Brazil}}
\author{E. L. Rodrigues}
\email{eduardor@cbpf.br}
\author{I. Dami\~ao Soares}
\email{ivano@cbpf.br}
\affiliation{{\it Centro Brasileiro de Pesquisas F\'{\i}sicas/MCT}\\
{\it R. Dr. Xavier Sigaud, 150. CEP 22290-180}\\
{\it Rio de Janeiro, RJ, Brazil}}

\date{\today}

\begin{abstract}
We present an alternative scheme of finding apparent horizons based on spectral methods applied to Robinson-Trautman spacetimes. We have considered distinct initial data such as representing the spheroids of matter and the head-on collision of two non-rotating black holes. The evolution of the apparent horizon is presented. We have obtained in some cases a mass gap between the final Bondi and apparent horizon masses, whose implications were briefly commented in the light of the thermodynamics of black holes. 
\end{abstract}

\maketitle

\section{Introduction}%

One of the most important problems in classical General Relativity is the evolution of apparent horizons. The apparent horizon \cite{AH} is defined as the outermost marginally trapped surface that can be located on each spacelike surface during the overall dynamics of the spacetime. According to cosmic censorship hypothesis there must exist outside the apparent horizon an event horizon \cite{AH}, and for this reason apparent horizons are the key structures that signalize the formation of black holes in gravitational collapse, as well play relevant role in the merging of black holes \cite{coal_bh}. Besides the apparent horizon another typical structure present in a spacetime that contains a black hole is the event horizon, but its determination depends on whole history of the spacetime due to the fact that an event horizon is the boundary that separates those null geodesics that reach infinity from those that not. In essence, while the event horizon is a global structure, the apparent horizon is local meaning that it can be determined at each instant. Therefore, the construction of apparent horizon finders is a crucial issue in numerical relativity that has received a great deal of attention in the last years \cite{AH_finders}. Basically, these codes are built to solve numerically the apparent horizon equation, which is a nonlinear elliptical equation, simultaneously with the numerical evolution of the spacetime.

In general most of the numerical strategies to solve the apparent horizon equation are based on the finite difference techniques. On the other hand, numerical codes based on spectral methods \cite{bonazzola,review_sm} have increased considerably in the last years mainly due to the economy of the computational resources to achieve a desired accuracy. In this direction we shall present here a simple and efficient numerical strategy using a convenient combination of Galerkin \cite{galerkin} and collocation methods \cite{boyd,canuto} to determine the evolution of the apparent horizon of Robinson-Trautman spacetimes \cite{rt}.

The Robinson-Trautman (RT) spacetimes are the simplest class of asymptotically flat geometries admitting gravitational waves with two interesting basic features: (a) a RT spacetime can be interpreted as describing the exterior geometry of a bounded or isolated system emitting gravitational waves; (b) for regular initial data the RT spacetimes evolve asymptotically towards the Schwarzschild black hole \cite{chru}. For the sake of completeness, the line element of the Robinson-Tratuman spacetimes can be conveniently written as

\begin{eqnarray}
ds^2 &=& \left(\lambda(u,\theta) - \frac{2 m_0}{r} + 2 r \frac{\dot{K}}{K}\right) d u^2 + 2 du dr - \nonumber \\
& & r^{2}K^{2}(u,\theta)(d \theta^{2}+\sin^{2}\theta d \varphi^{2}), \label{eq1}
\end{eqnarray}

\noindent where dot means derivative with respect to the null coordinate $u$, $r$ is the radial coordinate, $(\theta,\varphi)$ are the usual angular coordinates, and $m_0$ is an arbitrary constant. The Einstein equations can be cast in the following form

\begin{equation}
\label{eq2} \lambda(u,\theta)=\frac{1}{K^2}-\frac{K_{\theta \theta}}{K^3}+\frac{K_{\theta}^{2}}{K^4}-\frac{K_{\theta}}{K^3}\cot
\theta
\end{equation}
\begin{equation}
-6 m_{0}\frac{\dot{K}}{K}+\frac{(\lambda_{\theta} \sin
\theta)_{\theta}}{2 K^2 \sin \theta}=0. \label{eq3}
\end{equation}

\noindent Here the subscript $\theta$ denote derivative with respect to the angle $\theta$; the function $\lambda(u,\theta)$ is the Gaussian curvature of the surfaces $(u={\rm{const.}},r={\rm{const.}})$. The structure of the field equations is typical of a characteristic problem \cite{winicour}, in which the first equation is a hypersurface equation relating the functions $\lambda(u,\theta)$ and $K(u,\theta)$, whereas the second equation is the evolution equation. Accordingly, once the initial data $K(u_0,\theta)$ is prescribed on a given null surface $u=u_0$, the hypersurface equation fixes $\lambda(u_0,\theta)$, and the evolution equation determines $K(u,\theta)$ on the next null surface, and whole process repeats providing the evolution of the spacetime.

The only known analytical solutions of the RT field equations are the two forms of the Schwarzschild solution described by

\begin{eqnarray}
K = K_0 = \mathrm{constant}, \label{eq4}\\
\nonumber \\
K(\theta) = \frac{\bar{K_0}}{\cosh\gamma + \cos\theta \sinh\gamma}. \label{eq5}
\end{eqnarray}

\noindent The first reproduces the Schwarzschild black hole with mass $M_{BH} = m_0 K_0^3$, and the second expression a boosted black hole with constant velocity $v = \tanh \gamma$ with respect to an inertial observer at infinity. In this case the total mass-energy content is given by

\begin{equation}
M_{BH} = \frac{m_0 \bar{K_0}^3}{\sqrt{1-v^2}}. \label{eq6}
\end{equation}

\noindent Notice that this above expression corresponds to the total relativistic energy of a moving particle with rest mass $m_0 \bar{K_0}^3$ and velocity $v$.

This paper is divided as follows. In Section 2 we present the apparent horizon equation and the numerical strategy based on spectral methods adopted to solve it. In Section 3 we exhibit the numerical results that consists in testing the code along with the dynamics of the apparent horizon corresponding to initial data representing spheroids \cite{rt_radiation} and the collision of black holes \cite{rt_collision} in RT spacetimes. Finally, the final remarks are presented in Section 4.

\section{Solving the apparent horizon equation using spectral methods}%

As already mentioned, RT spacetimes have interesting features such as the asymptotic flatness and the presence of gravitational waves, which can be interpreted as arising from a bounded distribution of matter evolving towards a Schwarzschild black hole, and therefore indicating a simple example of non-spherical collapse. However, these geometries do not have a future apparent horizon characterized by the vanishing of the null expansion associated to outgoing future directed rays, but only past apparent horizon \cite{penrose,tod,chow_lun}. A past apparent horizon is the outermost boundary of past-trapped surfaces corresponding to that value of $u$; more precisely, consider a hypersurface $S$ defined by $S = r-V(u,\theta) = 0$, and in particular if $S$ is a marginally past-trapped 2-surface, the ingoing normal null vector $n_\alpha=\partial_\alpha S$ has vanishing divergence

\begin{equation}
\theta_{-} = n^\alpha_{;\alpha}=0. \label{eq7}
\end{equation}

\noindent From this equation it can be shown \cite{tod} that the function $V(u,\theta)$ satisfies the following equation at each slice $u=$ constant,

\begin{equation}
\frac{1}{\sin \theta}\,\left(\sin \theta \frac{V_\theta}{V}\right)_\theta - \lambda K^2 + \frac{2 m_0}{V}K^2 = 0, \label{eq8}
\end{equation}

\noindent which is known as the apparent horizon equation. The dynamics of the apparent horizon is obtained after solving this equation at each hypersurface $u=\mathrm{constant}$, where the function $K(u,\theta)$ is determined from the evolution equation (\ref{eq3}). There are few analytical results about the properties of past apparent horizons in RT spacetimes. Tod \cite{tod} has shown the validity of the isoperimetric inequality and the existence of a unique marginally past-trapped surface at each hypersurface $u$=constant.

The apparent horizon equation (\ref{eq8}) will be solved here using a numerical scheme based on a suitable combination of Galerkin and collocation methods in a similar way we have implemented to solve the field equations (\ref{eq2}) and (\ref{eq3}). For this reason we shall briefly outline our previous numerical scheme \cite{rt_prd,rt_ijmpc} for solving the field equations and, in the sequence, the procedure employed to integrate the apparent horizon equation.

According to Ref. \cite{rt_ijmpc} the first step is to establish the Galerkin expansion for the function $K(u,\theta)$,

\begin{equation}
K_a^2(u,x) = {\rm e}^{Q_a(u,x)}={\exp}\left(\sum_{k=0}^{N}\,b_k(u) P_k(x)\right),
\label{eq9}
\end{equation}

\noindent where the subscript $a$ indicates an approximation of the exact $K(u,x)$. The angular coordinate $\theta$ is replaced by $x=\cos \theta$, $N$ is the truncation order that indicates where the series stops, and the $N+1$ modes $b_k(u)$ are unknown functions of $u$ to be determined; the Legendre polynomials $P_k(x)$ were chosen as the basis or the trial functions. Next, an approximate expression for the function $\lambda(u,x)$ is obtained after substituting the above expansion into the constraint equation (\ref{eq2}), or

{\small
\begin{eqnarray}
\lambda_a(u,x)=\mathrm{e}^{-Q_a(u,x)}\left(1 + \sum_{k=0}^{N}\frac{k(k+1)}{2} b_k(u) P_k(x)\right).
\label{eq10}
\end{eqnarray}
}


These last two equations are substituted into Eq. (\ref{eq3}) to yield what is know as the residual equation associated to the evolution equation,

\begin{eqnarray}
\mathrm{Res}_K(u,x)=6\,m_0\,\sum_{k=0}^{N}\,\dot{b}_k(u) P_k(x) - \nonumber \\
{\rm e}^{-Q_a(u,x)}\,\Big[(1-x^2)\,\lambda_a^{\prime}\Big]^{\prime},
\label{eq11}
\end{eqnarray}

\noindent where prime denotes derivative with respect to $x$. Notice that the residual equation does not vanish exactly due to the adopted approximations for the functions $K(u,x)$ and $\lambda(u,x)$, but as we have shown it converges to zero as the truncation order $N$ is increased \cite{rt_ijmpc}. Following the Galerkin method, the projections of the residual equation with respect to a suitable set of test functions ${\Psi_n(x)}$ vanish, namely

\begin{eqnarray}
\left<\mathrm{Res}_K(u,x),\Psi_n(x)\right> = \int_{-1}^1\,{\rm Res}_K(u,x) \Psi_n(x)\,dx = 0,\nonumber \\
\label{eq12}
\end{eqnarray}

\noindent for $n=0,1,..N$. It means that the modes $b_j(u)$ are chosen in such a way that the residual equation is forced to be zero in an average sense \cite{finlayson}. 
Following the Galerkin method we have selected, for the above integrations, the test functions to be same as the trial functions, $\Psi_n(x)=P_n(x)$. At this point we have introduced an additional approximation for the exponential term given by

\begin{equation}
\exp(-Q_\mathrm{a}(u,x)) \approx \sum_{j=0}^{\bar{N}}\,c_j T_j(x), \label{eq13}
\end{equation}

\noindent where $T_j(x)$ is the Chebyshev polynomial of order $j$ and $\bar{N}$ indicates the number of modes $c_j$. Basically, the motivation behind this approximation is to allow rapid and direct integrations of the residual equation. As a consequence of the above expansion, the $\bar{N}+1$ modes $c_j$ are related to the $N+1$ modes $b_k$ by assuming that the projections of the residual equation $\mathrm{Res}_Q(u,x) = \exp(-Q_\mathrm{a}(u,x)) - \sum_{j=0}^{\bar{N}}\,c_j T_j(x)$ with respect to the test functions $\Psi_n(x)=\delta(x-x_n)$ vanish, where $x_0,x_1,..,x_{\bar{N}}$ are the collocation points associated to the Chebyshev polynomials.
The additional approximation (\ref{eq13}) is introduced into Eq. (\ref{eq12}) and after performing the $N+1$ integrals, a set of ordinary differential equations for the modes $b_k(u)$ arises. Therefore, evolving these equations means to determine the dynamics of RT spacetimes since the function $K(u,x)$ can be reconstructed at each $u$.

The past horizon equation (\ref{eq8}) will be solved at each time level $u$ by applying a similar combination of spectral methods as described previously. We have followed Ref. \cite{tod} and introduced an auxiliary function $F(u,x)$ by setting

\begin{equation}
V=2m_0\exp(-F), \label{eq14}
\end{equation}

\noindent in order to eliminate $m_0$ from the apparent horizon equation. A natural Galerkin expansion for $F(u,x)$ is given by

\begin{equation}
F_a(u,x) = \sum_{k=0}^M\,a_k(u) P_k(x), \label{eq15}
\end{equation}

\noindent where $M$ is the truncation order not necessarily the same as $N$ (see Eq. (\ref{eq9})). The apparent horizon equation is rewritten in function of $F(u,x)$, and after substituting the above expansion together with the approximate expressions for $K(u,x)$ and $\lambda(u,x)$ (Eqs. (\ref{eq9}) and (\ref{eq10})), we have obtained the residual equation associated to the apparent horizon equation

\begin{eqnarray}
\mathrm{Res}_{\mathrm{AH}}(u,x) &=& \left[(1-x^2) F_a^\prime\right]^\prime + 1 +\sum_{k=0}^{N}\,\frac{1}{2} k(k+1) \times \nonumber \\
& & b_k(u) P_k(x) - \exp(F_a+Q_a).
\label{eq16}
\end{eqnarray}


As we have described before, the next step is to impose that all projections of the residual equation (\ref{eq16}) with respect to each basis function, $P_n(x)$, $n=0,1,...,M$, must vanish. Schematically, we have

\begin{equation}
\left<\mathrm{Res}_{\rm AH},P_n(x)\right>=\int_{-1}^1\mathrm{Res}_{\rm AH}(u,x)P_n(x) = 0.
\label{eq17}
\end{equation}


\noindent Notice the presence of two exponential terms in the residual equation (\ref{eq16}) that can be reexpressed using additional approximations as,

\begin{eqnarray}
\exp(F_\mathrm{a}(u,x)) \approx \sum_{k=0}^{\bar{M}}\,\alpha_k T_k(x), \\
\exp(Q_\mathrm{a}(u,x)) \approx \sum_{k=0}^{\bar{M}}\,\beta_k T_k(x),
\end{eqnarray}

\noindent where $\alpha_k$ and $\beta_k$ are the modes associated to these new approximations, and $\bar{M}$ is the truncation order for both expansions. The projections of the corresponding residual equations, $\mathrm{Res}_{F} = \exp(F_{\mathrm{a}}(u,x)) - \sum_{k=0}^{\bar{M}}\,\alpha_k T_k(x)$ and $\mathrm{Res}_Q = \exp(Q_{\mathrm{a}}(u,x)) - \sum_{k=0}^{\bar{M}}\,\beta_k T_k(x)$, with respect to the test functions ${\delta(x-x_n)}$, with $x_n=0,1,..,\bar{M}$ being the collocation points of Chebyshev polynomials, are forced to vanish. Consequently, two sets of $\bar{M}+1$ algebraic equations relating the modes $(\alpha_k,\beta_k)$ with $(a_j,b_k)$, respectively, are generated. These approximate expressions are then inserted into Eq. (\ref{eq17}), yielding

{\small
\begin{eqnarray}
& & \left<\mathrm{Res}_{\rm AH},P_n(x)\right>=\int_{-1}^1\,\{\left[(1-x^2) F_a^\prime\right]^\prime + 1 + \frac{1}{2} \times \nonumber \\
& & \sum_{k=0}^{N}\,k(k+1)b_k(u) P_k(x) - \sum_{k,j=0}^{\bar{M}}\,\alpha_k(u)\beta_j(u)T_k(x) \times \nonumber \\
& & T_j(x)\}P_n(x) = 0.
\end{eqnarray}}

\noindent After performing the above integrals, a set of $M+1$ algebraic equations of the type $f_k(a_j,b_j,\alpha_i,\beta_i)=0$ is obtained. Since we can express the modes $\alpha_k$ and $\beta_k$ in terms of $a_j$ and $b_j$, and the modes $b_k$ are known at each $u$, we can solve, in principle, this set of algebraic equations for the modes $a_k$, and therefore determining the apparent horizon as described by Eq. (\ref{eq15}).



\section{Numerical results}%

In this section, we present the numerical tests of our code as well the results about the dynamics of apparent horizons in RT spacetimes. We first need to specify the initial data $K(u=0,x)$ that determine the initial values for the $N+1$ modes $b_k(0)$ through

\begin{equation}
b_j(0)=\frac{2 \left<\ln K(0,x),P_j\right>}{\left<P_j,P_j\right>}.
\label{eq20}
\end{equation}

\noindent We are going to consider two initial data in our numerical experiments. The first represents the exterior spacetime of a homogeneous oblate spheroid described by \cite{rt_radiation}

\begin{equation}
K(0,x)=\Big[1+\frac{B_0}{2}\Big(\alpha(\zeta_0)+\frac{1}{2}\beta(\zeta_0)P_2(x)\Big)\Big]^2,
\label{eq21}
\end{equation}

\noindent where $\zeta_0$ and $B_0$ are free parameters and $\alpha(\zeta_0)=\arctan(1/\zeta_0)$,  $\beta(\zeta_0)=(1+3\zeta_0^2)\arctan(1/\zeta_0)-3\zeta_0$. There is a clear astrophysical motivation for such a family of initial data as pointed out in the works on the axisymmetric gravitational collapse of oblate gas spheroids satisfying the Vlasov equation either in Newtonian theory \cite{lin}, as well in its relativistic generalization \cite{shapiro}; and also connected with the efficiency of the emission of gravitational waves \cite{eardley_spheroids}. The second initial data family describe two initially boosted Schwarzschild black holes with opposite velocities $v=\tanh \eta_0$ \cite{rt_collision} in which

\begin{eqnarray}
K(0,x)=\Big( \frac{A_{1}}{\sqrt{\cosh \eta_0-x\sinh \eta_0}} + \nonumber \\
+ \frac{A_{2}}{\sqrt{\cosh \eta_0+x\sinh \eta_0}} \Big)^2,
\label{eq22}
\end{eqnarray}

\noindent where $A_{1}$ and $A_{2}$ are arbitrary positive constants associated to the mass of each black hole.

We first exhibit an important test for the spectral code used to integrate numerically the field equations (\ref{eq2}) and (\ref{eq3}). In spite of non-stationary analytical solutions of the field equations are not known (unless in the linear regime), there exists a conserved quantity


\begin{equation}
I_0 = \int_{-1}^{1}\,K^2(u,x) dx,
\end{equation}

\noindent  which is derived from the field equations and interpreted as the area of the fundamental 2-sphere spanned by $(\theta,\phi)$.  The conservation of $I_0$ can be deduced after multiplying Eq. (\ref{eq3}) by $K^2$ and integrating in the angular domain. Then, by specifying the initial data $K(0,x)$, the initial value of $I_0$ will be fixed and must be kept constant until the asymptotic state is achieved. In order to test if the numerically generated solution is accurate, we have evaluated the relative error between the numerical and exact values of $I_0$, $\sigma = |I_0-I_{\mathrm{numer}}|/I_0$, whose result is shown in Fig. 1 where we have included the influence of increasing truncation orders, or $N=7,9,13$ (cf. Eq. (\ref{eq9})). As it can be observed, the conservation of $I_0$ is attained to about $10^{-8}\%$ accuracy for the smallest truncation order $N=7$, and about $10^{-12}\%$ accuracy for $N=13$. Therefore, this result is a vivid proof of the accuracy of the numerical evolution scheme despite those other tests described in Ref. \cite{rt_ijmpc}.


\begin{figure}[ht]
\vspace{-.5cm}
\rotatebox{270}{\includegraphics*[scale=0.27]{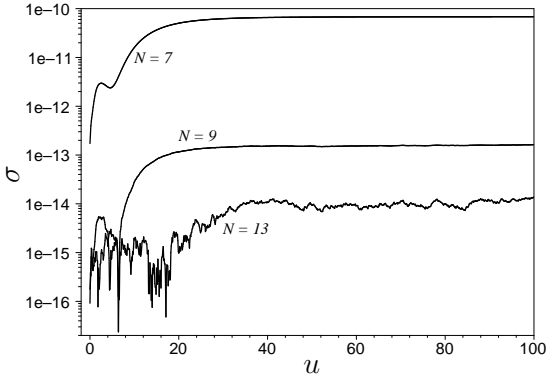}}
\caption{Evolution of the relative error $\sigma$ between $I_0$ and $I_{\mathrm{numer}}$. We have considered the initial data that represent two initially boosted Schwarzschild black holes (Eq. (\ref{eq22})) with $A_1=1.0$, $A_2=0.2$ and $\eta_0=0.3$, as well distinct truncation orders $N=7,9,13$ associated to the Galerkin expansion of $K(u,x)$ given by Eq. (\ref{eq9}).}
\end{figure}

We now proceed with the numerical tests of the algorithm used to solve the apparent horizon equation (\ref{eq8}) at each time level $u$. Two steps will be needed. The first is to evaluate the time evolution of all modes $b_k(u)$ by integrating the dynamical system resulting from (\ref{eq12}). In the second step these modes calculated at each $u$ are inserted into the system of $M+1$ algebraic equations derived from (20) and solve them to obtain the corresponding modes  $a_k(u)$ that describe the apparent horizon through the function $V(u,x)$ (cf. Eqs. (\ref{eq14}) and (\ref{eq15})). In this way, the evolution of the apparent horizon is obtained until the stationary solution is attained. As a matter of fact, as an important piece of evidence of the accuracy of our numerical scheme, we have plotted in Fig. 2 the modulus of the residual equation (\ref{eq16}) evaluated at the initial instant $u=0$ for both initial data families (\ref{eq21}) and (\ref{eq22}) taking into account distinct values of the truncation orders $N$ and $M$ associated to the functions $K(u,x)$ and $V(u,x)$ (cf. Eqs. (\ref{eq9}) and (\ref{eq15})). According to these plots the residual equation approach to zero as a consequence of increasing the truncation orders under consideration.

\begin{figure}[ht]
\rotatebox{270}{\includegraphics*[scale=0.28]{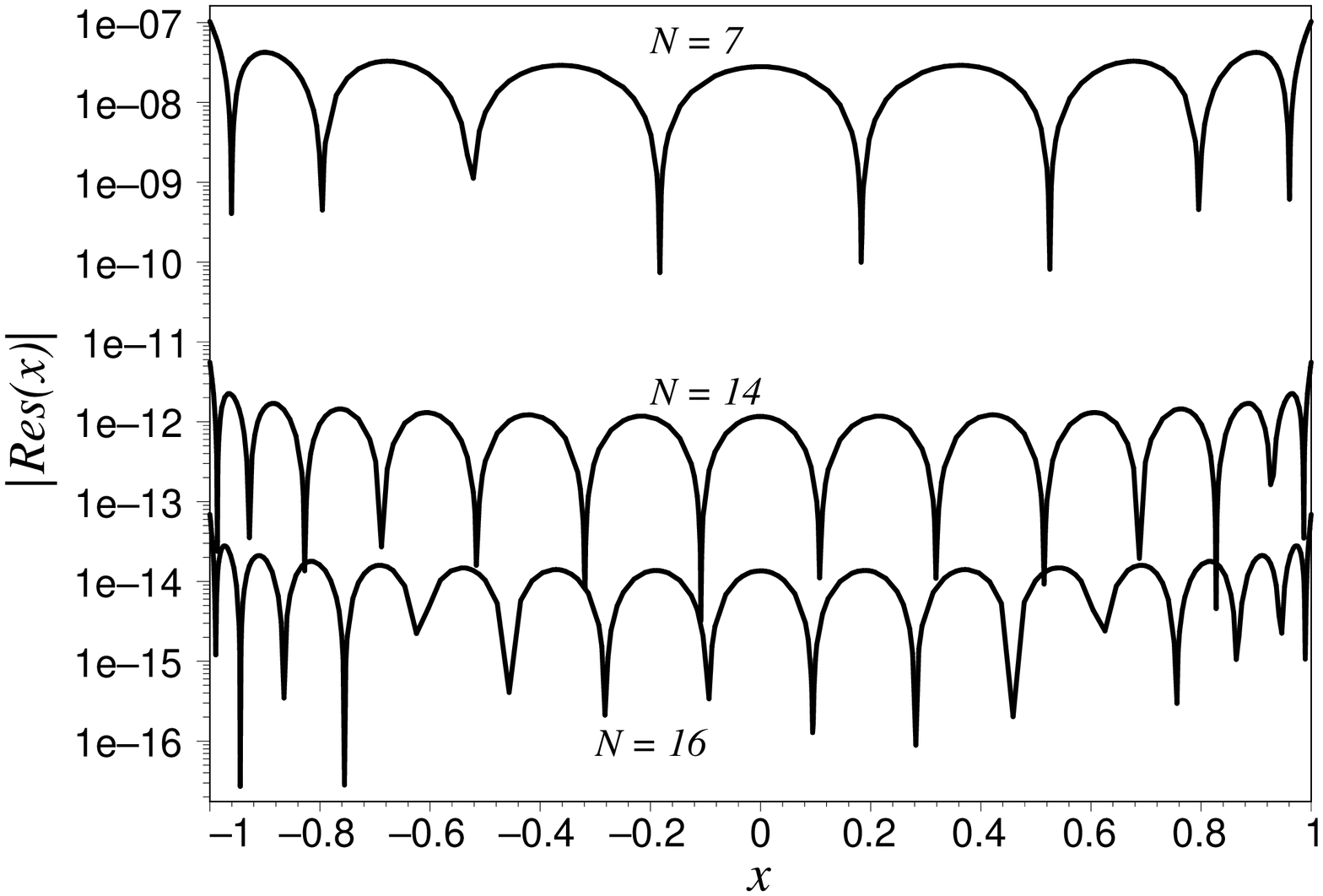}}
\rotatebox{270}{\includegraphics*[scale=0.28]{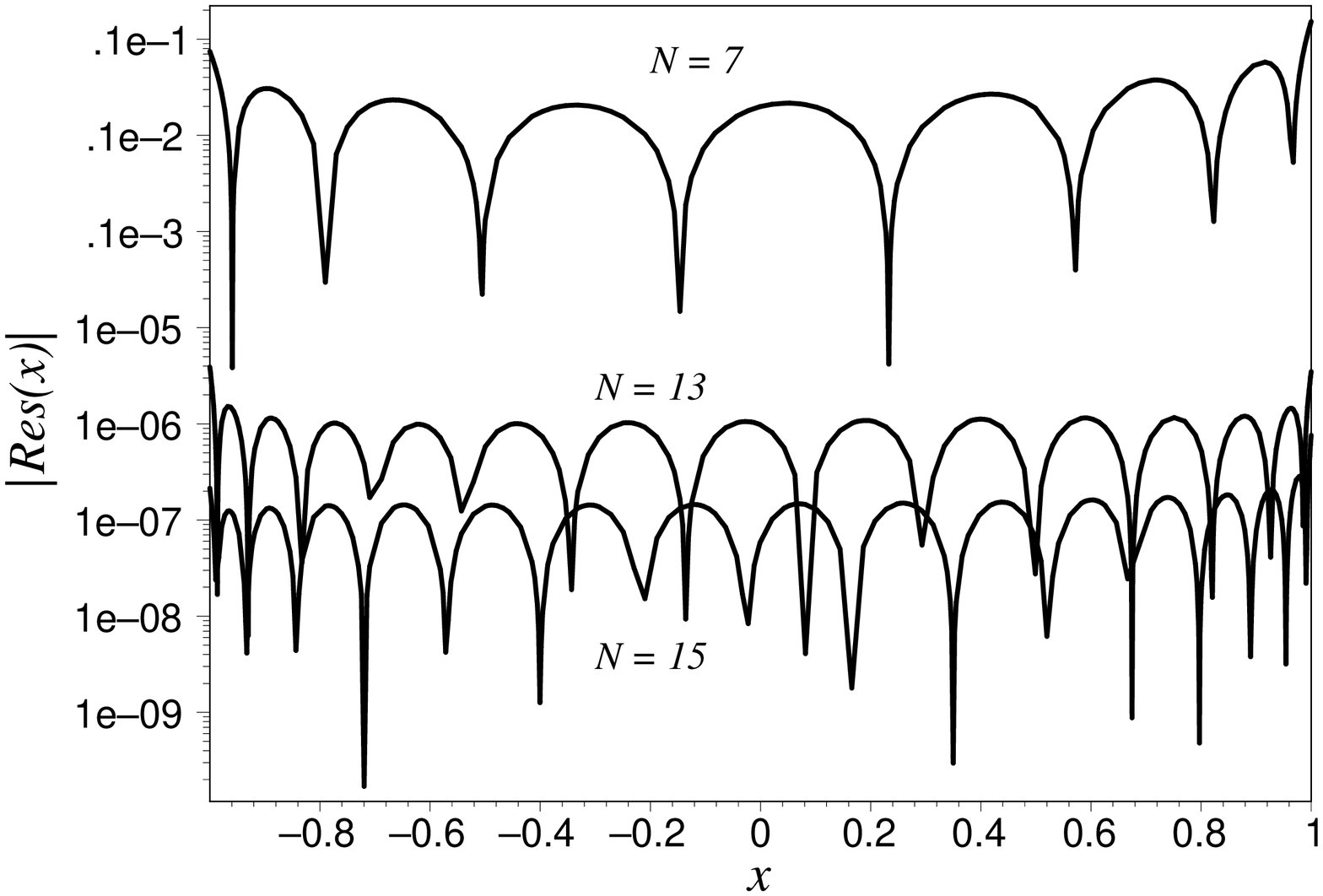}}
\caption{Log-linear plots of the residual equation associated to the apparent horizon equation (\ref{eq16}) showing the effect of increasing the truncation orders $N$ as indicated above and $M=N$ in each case. Notice the decrease of the residual equation as more terms are included in the Galerkin expansion. The initial data are the oblate spheroid given by (\ref{eq21}) with $\zeta_0=0.1$ and $B_0=1$ (first graph), and the two initially boosted black holes (Eq. (\ref{eq22})) in which $A_1=1.0$, $A_2=0.1$ and $\eta_0=0.3$.}
\end{figure}

A more enlightening experiment for depicting the convergence of the code is to exhibit the evolution of the $L_2$ norm corresponding to the residual equation (\ref{eq16}) given by

\begin{equation}
L_2 = \sqrt{\frac{1}{2}\int^{1}_{-1}{{\rm Res_{\mathrm{AH}}}(u,x)^{2}dx}},
\label{eq23}
\end{equation}

\noindent considering again distinct values of the truncation orders $N$ and $M$.  From Fig. 3 it can be seen that the norm evaluated at $u=0.4$ decays exponentially if the truncation order $N$ is increased, which demonstrates the expected geometric convergence typical of spectral methods. In Fig. 4 the full evolution of $L_2$ is presented for increasing truncation orders $M,N$, and as expected it is noticed a rapid decreased of the norm until reaching to the value considered zero up to our numerical precision; also when the truncation order is increased, less time is necessary to reach to that value.


The evolution of the apparent horizon is illustrate by a sequence of polar plots of $r=V(u,x)$ and depicted in Fig. 5. We started at $u=0$ with the oblate spheroid initial data (\ref{eq21}) and several plots are shown in subsequent instants until $u_f = 500$ where a circle is formed. Indeed, this is a consequence from the fact that the asymptotic state is the Schwarzschild configuration characterized by $K=\mathrm{constant}$ according to Eq. (\ref{eq8}) which implies also in $V=\mathrm{constant}$.

\begin{figure}[ht]
\rotatebox{0}{\includegraphics*[scale=0.25]{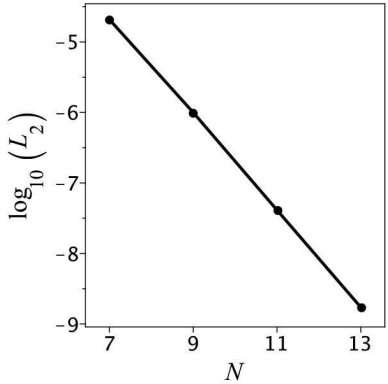}}
\vspace{-2.5cm}
\caption{The $L_2$ norm of the residual equation (\ref{eq23}) evaluated at $u=0.4$ and for distinct truncation orders. Notice that the exponential decay of the norm with the increase of the truncation order is typical of spectral methods. Here we have used the two boosted black holes (\ref{eq22}) with $A_1=1.0$, $A_2=0.15$ and $\eta_0=0.4$ as the initial data.}
\end{figure}

\begin{figure}[ht]
\rotatebox{270}{\includegraphics*[height=7cm,width=5.5cm]{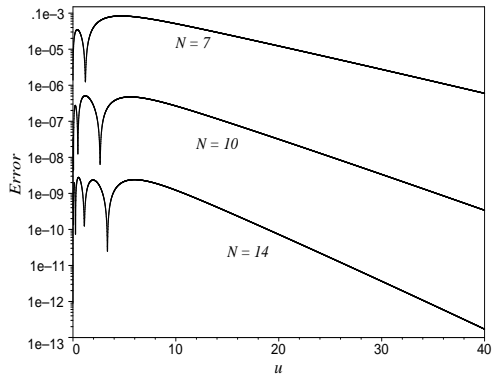}}
\caption{Behavior of the $L_2$ norm of the apparent horizon equation (cf. Eq. (\ref{eq23})) for the oblate spheroid (\ref{eq21}) with $\zeta_0=0.1$ and $B_0=0.3$ taking into account distinct truncation orders $N$ and with $M=N$ in each case. Again as far as $N$ is increased more rapidly the norm tend to zero.}
\end{figure}

\begin{figure}[htb]
\vspace{-1cm}
\rotatebox{0}{\includegraphics*[scale=0.3]{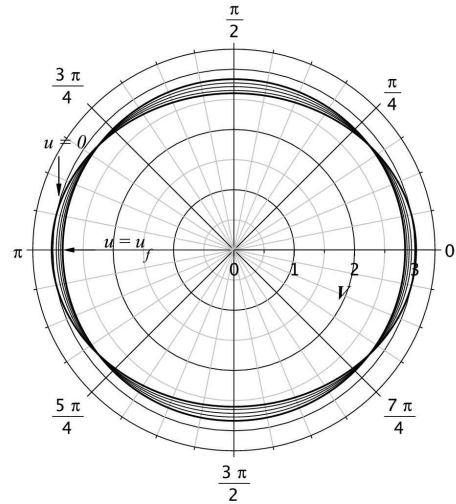}}
\vspace{-2cm}
\caption{Dynamics of all modes $a_k(u)$ through a sequence of polar plots of $2m_0\exp(-F_{\mathrm{a}}(u,x))$ starting from the oblate spheroid initial data ($\zeta_0=0.1$ and $B_0 = 0.4$) at $u_0=0$ as indicated, and for several times until the Schwarzschild final configuration (represented by a circle) is formed at approximately $u_f =500$.}
\end{figure}

An interesting application of our code is to follow the behavior of the apparent horizon mass that is basically the amount of mass enclosed by the apparent horizon. It is worth of mentioning that the apparent horizon mass has thermodynamical properties similar to those associated to black holes \cite{chow_lun}. In the case of RT spacetimes the past apparent horizon can only decrease in area, and therefore its mass decreases, contrary to the monotonic increase of the future apparent horizon area. The apparent horizon area $S_{AH}$ is evaluated through the following expression

\begin{eqnarray}
S_{AH} &=& 2 \pi \int_{-1}^1\,r^2 K^2(u,x) dx = \nonumber \\
&=& 8 \pi m_0^2 \int_{-1}^1\,\mathrm{e}^{-2F(u,x)} K^2(u,x) dx,\label{eq24}
\end{eqnarray}

\noindent where $r=V(u,x)=2m_0 \mathrm{e}^{-2F(u,x)}$ describes the apparent horizon (cf. Eq. (\ref{eq14})), and the apparent horizon mass is expressed as

\begin{equation}
M_{AH} = \sqrt{\frac{S_{AH}}{16\pi}}. \label{eq25}
\end{equation}

In Fig. 6 we present the evolution of the apparent horizon mass and the Bondi mass \cite{rt_prd2,kramer}

\begin{equation}
M_{B} = \frac{1}{2}m_0\,\int_{-1}^1 K^3(u,x) dx, \label{eq26}
\end{equation}

\noindent for the first family of initial data (\ref{eq21}). According to Ref. \cite{rt_radiation} the asymptotic configuration is the Schwarzschild black hole whose final mass assumes the value $M_{BH}=m_0K_0^3$, where $K_0=\lim_{u\rightarrow \infty}\,K(u,x)$. This amount is smaller than the mass associated to the initial data since part of it is extracted by gravitational waves \cite{bondi} during the evolution of the spacetime, and consequently producing a monotonic decrease of the Bondi mass as shown in Fig. 6. Nonetheless, the decay of the apparent horizon mass is due to the decrease in area of the past apparent horizon as we have mentioned before. Notice also that both final values of the apparent horizon and Bondi masses coincide. As a matter of fact, this result is expected from the asymptotic solution of the apparent horizon equation  $V_{\mathrm{asympt}}=2m_0K_0^2$, and together with the expression for the apparent horizon mass (\ref{eq25}) one can arrive at $M_{AH}=M_{BH}=m_0K_0^3$.

\begin{figure}[h]
\vspace{-0.7cm}
\includegraphics*[scale=0.35]{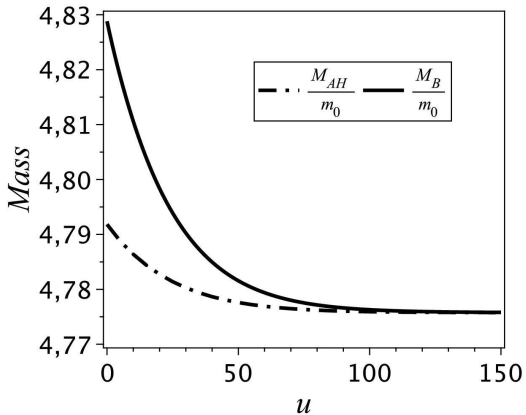}
\vspace{-3.5cm}
\caption{Numerical evolution of the Bondi and apparent horizon masses in units of $m_0$ attesting their monotonic decay until the final configuration identified as the Schwarzschild black hole in which both values coincides. The initial data is the oblate spheroid with $\zeta_0=0.1$ and $B_0=0.4$.}
\end{figure}

The final task is to consider the second family of initial data (\ref{eq22}) which represent the head-on collision of two Schwarzschild black holes and a generalization of the initial data (\ref{eq21}) that describe the exterior of an inhomogeneous oblate spheroid \cite{rt_radiation}. The common feature of both initial data is that the asymptotic configuration will be a boosted black hole \cite{rt_collision,rt_radiation} described by

\begin{equation}
\lim_{u \rightarrow \infty}\,K(u,x) = \frac{\bar{K_0}}{\cosh \mu + x \sinh \mu}, \label{eq27}
\end{equation}

\noindent where the values of $\bar{K_0}$ and the boost parameter $\mu$ are fixed by the numerical solution of the RT equation (see Ref. \cite{rt_collision} for details), and the final Bondi mass is given by Eq. (\ref{eq6}). It is worth mentioning that the imbalance in momentum of the initial gravitational wave distribution is responsible for the boost of the resulting black hole. In Figs. 7(a) and 7(b) we observe again the monotonic decay of both $M_{AH}$ and $M_B$ with the retarded time $u$, but there is a gap between their asymptotic values. In order to understand the origin of this gap, we have noticed that according to the numerical experiments the asymptotic solution of the apparent horizon equation (\ref{eq8}) is the same as the previous case, $V_{\mathrm{asympt}} = 2 m_0 \bar{K_0}^2$, in spite of $K_{\mathrm{asympt}}$ not being a constant (cf. Eq. (\ref{eq26})). Therefore, the final value of the apparent horizon mass can be evaluated from  Eq. (\ref{eq25}) (note that in this situation $K=K(x)$), and whose result is $M_{AH}=m_0 \bar{K_0}^3$. In fact, this value is exactly the rest mass of the boosted black hole and consequently the gap observed in both graphs of Fig. 7 is due to the kinetic energy of the resulting black hole.

The above results can be interpreted in the light of the so called
First Law of Black Hole Thermodynamics \cite{beken}-\cite{wald}. The final mass configurations
displayed in each of Figs. 7 can actually be interpreted as two static black holes
boosted with respect to the each other,  namely, they are connected by a $K$-transformation
of the BMS group \cite{bondi} corresponding to a boost along the $z$-axis, as given by Eq. (\ref{eq27}).
This is the origin of the gap in Figs. 7, which has the value $M_{\rm B}/M_{AH}={\rm cosh} \mu$,
where $\mu$ is the boost parameter of the $K$-transformation specified in (\ref{eq27}).
Now the entropy of each final black hole, considered as a thermodynamical system in
equilibrium, is defined as proportional to the area $\mathcal{A}$ of its event horizon and
is invariant by a $K$-transformation as can be easily verified. This should be expected since
an eventual possible definition of the BH entropy by a counting of its microscopic states
could not depend, in principle, on the state of motion of stationary black holes relative
to inertial frames at infinity. Therefore we have $\delta \Big({\mathcal A}/4 \pi G \Big) = \delta M_{B}/T_{B}=\delta M_{AH}/T_{AH}$ so that this gap also defines the temperature transformation
$T_{B} \rightarrow T_{B}~{\rm cosh} \mu$ between the two inertial rest frames of the BHs.

\begin{figure}[ht]
\rotatebox{90}{\includegraphics*[scale=0.29]{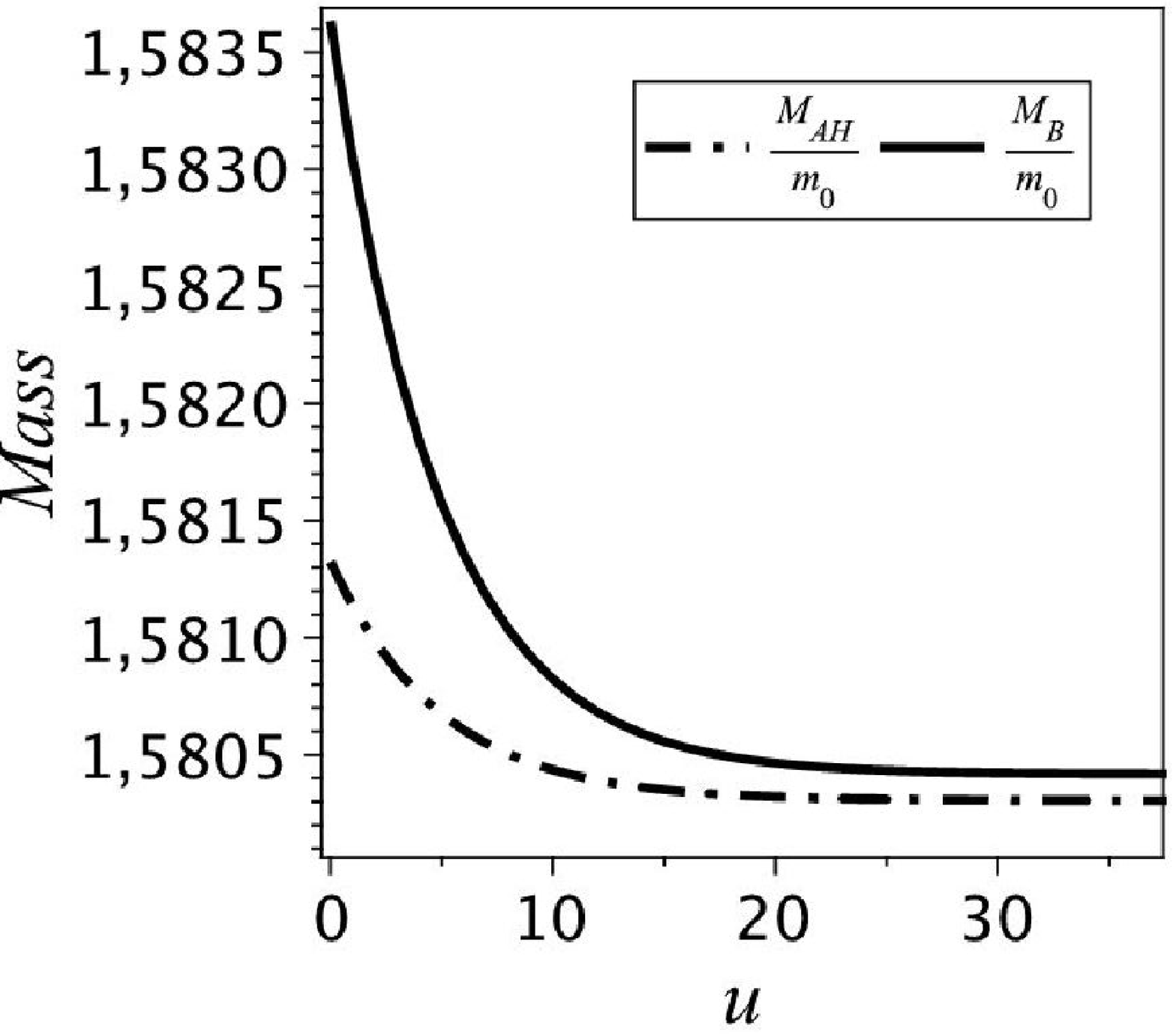}}
\rotatebox{90}{\includegraphics*[scale=0.29]{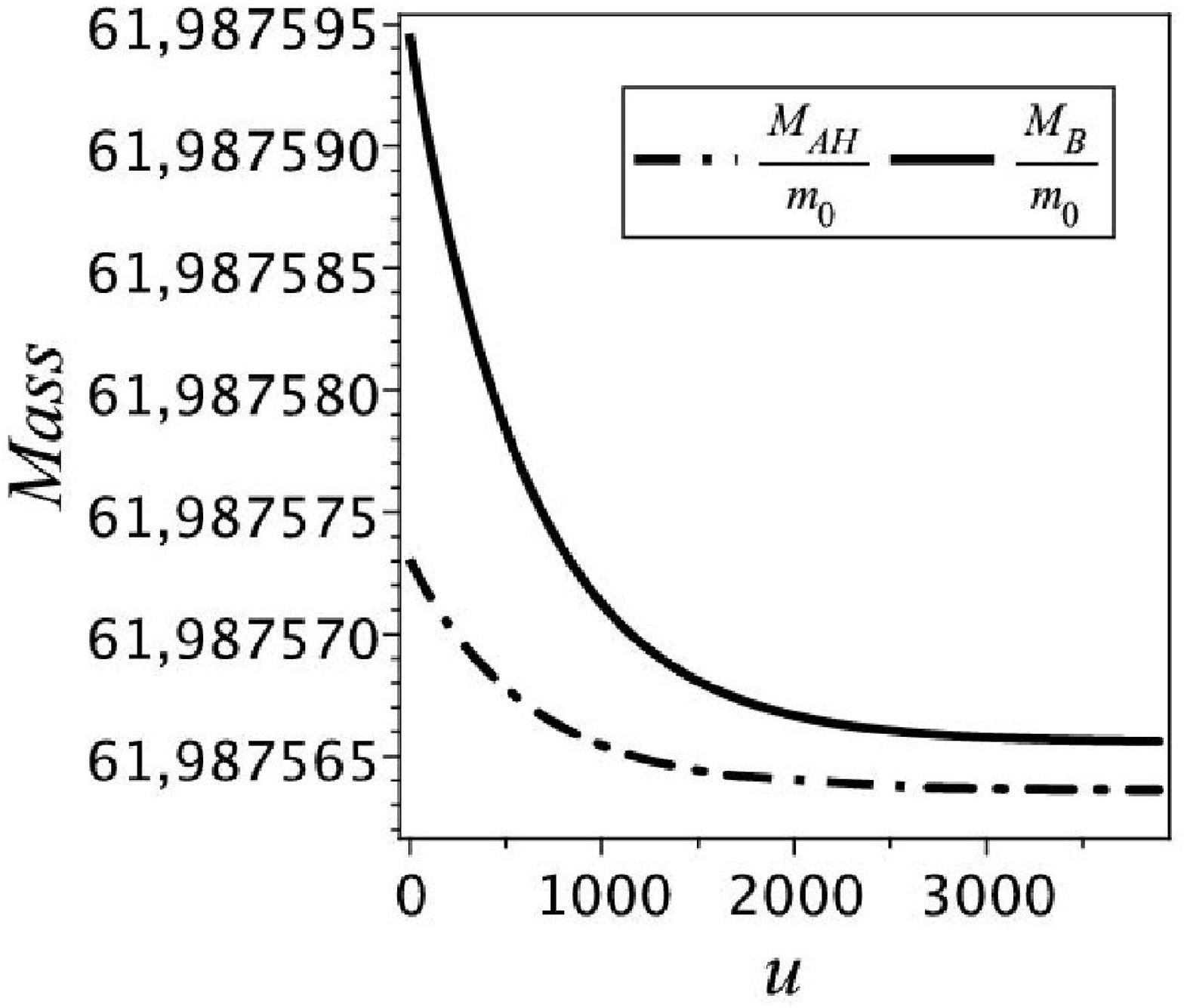}}
\caption{Numerical decays of Bondi and apparent horizon masses in units of $m_0$. Here the final configuration is a boosted black hole and both values of mass do not coincide asymptotically. The Bondi mass given by Eq. (\ref{eq6}) gather the rest and kinetic energy of the hole, whereas the apparent horizon mass $M_{AH}$ is a measure of the rest mass. We have considered in the first pannel a nonhomogeneous oblate spheroid initial data \cite{rt_radiation} given by $K(0,x)=\Big[1+0.05\Big(\alpha(\zeta_0)+\frac{1}{2}\beta(\zeta_0)P_2(x)\Big)+0.05 \exp (x-0.3)^2\Big]^2$, $\zeta_0=0.1$, whereas in the second pannel the head-on collision of two boosted black holes with $\eta_0=0.05$, $A_1=1.0$ and $A_2=0.99$.}
\end{figure}

\section{Final considerations}%

In this paper we have implemented and tested a numerical scheme based on a combination of Galerkin and pseudo-spectral methods to solve the apparent horizon equation in RT spacetimes. This a direct extension of the previous algorithm  \cite{rt_ijmpc} used to integrate the field equations (\ref{eq2}) and (\ref{eq3}). The apparent horizon equation is reduced to a set of nonlinear algebraic equations for the modes $a_k$ and whose solution at each instant determines the apparent horizon described by Eq. (\ref{eq8}). The applications have consisted in solving the apparent horizon corresponding to initial data describing the exterior fields of oblate spheroids and the collision of two Schwarzschild black holes.

We have performed numerical tests that strongly indicate the convergence and accuracy of the code. In our numerical experiments two initial data families in RT spacetimes were considered: the first represents the gravitational field outside a oblate spheroid while the second two initially boosted Schwarzschild black holes with opposed velocities. We have confirmed that the Bondi mass $M_B$ decreases monotonically as the result of the mass extraction due to the gravitational waves, until a asymptotic value that coincides with the total mass of the resulting Schwarzschild black hole. The apparent horizon mass $M_{AH}$ also decreases with respect to $u$ in face of the decreased in area which is expected in the case of past apparent horizon. By considering the first initial data, the asymptotic values of $M_B$ and $M_{AH}$ coincide to the value of the total mass of the resulting Schwarzschild black hole. In this case the apparent horizon mass is exactly the mass enclosed by the event horizon. On the other hand, if we take into account the second initial data a gap between the asymptotic values of both masses is observed similarly as noticed by Chow and Lun \cite{chow_lun}. The origin of the gap is associated to the final configuration identified as a boosted Schwarzschild black hole for which the Bondi mass is the total mass-energy content that includes the rest mass $m_0\bar{K}_0^3$ plus the kinetic energy, whereas the final apparent horizon mass is the rest mass. 

Finally, in spite of RT spacetimes being the simplest asymptotically flat radiating geometries, they can be potentially used as simple but useful theoretical laboratories to study relevant features of the bounded sources emitting gravitational waves (see Refs. \cite{rt_radiation}, \cite{rt_collision}, \cite{rt_prd} and \cite{rt_bremss}), and also to test new numerical schemes such the one we have implemented here. The natural step in our research is to examine the evolution of apparent horizons in the case of general RT spacetimes, and also in more realistic frameworks such as for spacetimes with Brill waves.

\begin{acknowledgments}
The authors acknowledge the financial support of the Brazilian agencies CNPq and FAPERJ.
\end{acknowledgments}


\begin{thebibliography}{99}

\bibitem{AH} S. W. Hawking and G. F. R. Ellis, \textit{The Large Scale Structure of Spacetime} (Cambridge University Press, Cambridge, England, 1973).

\bibitem{coal_bh} Frans Pretorius, Phys. Rev. Lett. \textbf{95}, 121101 (2005).

\bibitem{AH_finders} Jonathan Thornburg, \textit{Event and Apparent Horizon Finders for 3+1 Numerical Relativity},
Living Rev. Relativity 10,  (2007),  3. http://www.livingreviews.org/lrr-2007-3

\bibitem{bonazzola} S. Bonazzola, E. Gourgoulhon and J. A. Marck, J. Comput. Appl. Math. \textbf{109}, 433 (1999).

\bibitem{review_sm} Philippe Grandclément and Jérôme Novak, \textit{Spectral Methods for Numerical Relativity},
Living Rev. Relativity 12,  (2009),  1. http://www.livingreviews.org/lrr-2009-1

\bibitem{galerkin}  P. Holmes. John L. Lumley and Gal Berkooz, {\it Turbulence, Coherent Structures, Dynamical Systems and Symmetry}, Cambridge University Press (Cambridge, 1998).

\bibitem{boyd} J. P. Boyd, {\it Chebyshev and Fourier Spectral Methods}, Dover (2001).

\bibitem{canuto} C. Canuto, M. Y. Hussaini, A. Quarteroni and T. A. Zang, {\it Spectral Methods, Fundamentals in Single Domains}, Springer (2006).

\bibitem{rt} I. Robinson and A. Trautman, Phys. Rev. Lett. \textbf{4}, 431 (1960); Proc. Roy. Soc. A\textbf{265}, 463 (1962).

\bibitem{chru} P. Chrusciel, Commun. Math. Phys. 137, 289 (1991); Proc. Roy. Soc. London \textbf{436}, 299 (1992); P. Chrusciel and D. B. Singleton, Commun. Math. Phys. \textbf{147}, 137 (1992).

\bibitem{winicour} Jeffrey Winicour, \textit{Characteristic Evolution and Matching}, Living Rev. Relativity 8,  (2005),  10. http://www.livingreviews.org/lrr-2005-10

\bibitem{rt_radiation} H. P. de Oliveira and E. L. Rodrigues, Class. Quantum Grav. \textbf{25}, p. 205020 (2008).

\bibitem{rt_collision} R. Aranha, H. P. de Oliveira, I. D. Soares and E. V. Tonini, Int. J. Mod. Phys. D, \textbf{17}, 1 (2008)

\bibitem{penrose} R. Penrose, Ann. NY Acad. Sci. \textbf{224}, 115 (1973)

\bibitem{tod} K. P. Tod, Class. Quantum Grav., \textbf{6}, 1159 (1989).

\bibitem{chow_lun} E. W. Chow and A. W. Lun, \textit{Apparent Horizons in Vacuum Robinson-Trautman
Spacetimes}, preprint gr-qc 9503065.

\bibitem{rt_prd} H. P. de Oliveira and I. Dami\~ao Soares, Phys. Rev. D\textbf{70}, 084041 (2004).

\bibitem{rt_ijmpc} H. P. de Oliveira, E. L. Rodrigues, I. Dami\~ao Soares and E. V. Tonini, Int. J. Mod. Phys. C, \textbf{18}, 1853 (2007).    

\bibitem{finlayson} Bruce A. Finlayson, \textit{The Method of Weighted Residuals and Variational Principles}, Academic Press (1972).

\bibitem{lin} C. C. Lin, L. Mestel and F. H. Shu, Astrophys. J. \textbf{142}, 1431 (1965).

\bibitem{shapiro} Stuart L. Shapiro and Saul L. Teukolsky, Phys. Rev. Lett., \textbf{66}, 994 (1991); Phys. Rev. D\textbf{45}, 2006 (1992).

\bibitem{eardley_spheroids} D. M. Eardley, Phys. Rev. D \textbf{12}, 3072 (1975).

\bibitem{kramer} U. von der G\"onna and D. Kramer, Class. Quant. Grav. \textbf{15}, 215 (1998).

\bibitem{rt_prd2} H. P. de Oliveira and I. Dami\~ao Soares, Phys. Rev. D \textbf{71}, 124034 (2005).

\bibitem{bondi} H. Bondi, M. G. J. van der Berg and A. W. K. Metzner, Proc. R. Soc. London Ser. A\textbf{269}, 21 (1962); R. K. Sachs, Phys. Rev. \textbf{128}, 2851 (1962).

\bibitem{beken} J. D. Bekenstein, Phys. Rev. D \textbf{7}, 2333 (1973).

\bibitem{wald} R. M. Wald, {\it General Relativity} (University of Chicago Press, Chicago, 1984).

\bibitem{rt_bremss} H. P. de Oliveira, I. Dami\~ao Soares and E. V. Tonini, Phys. Rev. D \textbf{78}, 044017 (2008). 

\end{thebibliography}
\end{document}